# Charge-order-assisted topological surface states and flat bands in the kagome superconductor $CsV_3Sb_5$

Yong Hu[1,2,#], Samuel M. L. Teicher[3,#], Brenden R. Ortiz[3,#], Yang Luo[1], Shuting Peng[1], Linwei Huai[1], J. Z. Ma[4], N. C. Plumb[2], Stephen D. Wilson[3,*], J.-F. He[1,*] and M. Shi[2,*]

[1]*Hefei National Laboratory for Physical Sciences at the Microscale, Department of Physics and CAS Key Laboratory of Strongly-coupled Quantum Matter Physics, University of Science and Technology of China, Hefei, Anhui 230026, China*
[2]*Swiss Light Source, Paul Scherrer Institute, CH-5232 Villigen PSI, Switzerland*
[3]*Materials Department and California Nanosystems Institute, University of California Santa Barbara, Santa Barbara, California 93106, USA*
[4]*Department of Physics, City University of Hong Kong, Kowloon, Hong Kong*

[#]These authors contributed equally to this work.
*To whom correspondence should be addressed:
J.-F.H.(jfhe@ustc.edu.cn), M.S.(ming.shi@psi.ch), S.D.W.(stephendwilson@ucsb.edu).

**The diversity of emergent phenomena in quantum materials often arises from the interplay between different physical energy scales or broken symmetries [1-9]. Cooperative interactions among them are rare; however, when they do occur, they often stabilize fundamentally new ground states or phase behaviors [10-12]. For instance, a pair density wave can form when the superconducting order parameter borrows spatial periodical variation from charge order [13-16]; a topological superconductor can arise when topologically nontrivial electronic states proximitize with or participate in the formation of the superconducting condensate [1,4,17-21]. Here, we report spectroscopic evidence for a unique synergy of topology and correlation effects in the kagome superconductor $CsV_3Sb_5$ - one where topologically nontrivial surface states are pushed below the Fermi energy ($E_F$) by charge order, making the topological physics active near $E_F$ upon entering the superconducting state. Flat bands are observed, indicating that electron correlation effects are also at play in this system. Our results reveal the peculiar electronic structure of $CsV_3Sb_5$, which holds the potential for realizing Majorana zero modes [22] and anomalous superconducting**

**states [16,23-25] in kagome lattices. They also establish CsV$_3$Sb$_5$ as a unique platform for exploring the cooperation between the charge order, topology, correlation effects and superconductivity.**

Exotic quantum phenomena may appear in material systems with multiple orders or phases, where the mutual interactions can give rise to new physics beyond that of each component [1-12]. The newly discovered kagome-lattice CsV$_3$Sb$_5$ is such a system. It was the first member of the AV$_3$Sb$_5$ (A= K, Rb, Cs) shown to host the unique combination of a superconducting ground state (T$_c$=2.5 K) and a charge density wave (CDW) order below T$_{CDW}$=94 K (Fig. 1a) [26]. Nontrivial topology and strong electron correlation effects from the unique geometry of the kagome lattice are also expected, which would exhibit Dirac fermions and flat bands, respectively (Fig. 1b) [27-31]. In particular, topologically nontrivial surface states (TSSs) are theoretically suggested in the band structure of this material [26]. However, the calculated TSSs in CsV$_3$Sb$_5$ are above the Fermi energy (E$_F$) [26], making interactions with superconductivity unlikely. In this study, by using angle-resolved photoemission spectroscopy (ARPES), we find a cooperative action between the charge order and topological states in CsV$_3$Sb$_5$. The TSSs are pushed below E$_F$ by the electronic band reconstruction driven by the charge order, which activates the topological states for electron conduction in the superconducting state at low temperature. We also observe flat bands in the material, indicating that electron correlation effects need to be considered.

CsV$_3$Sb$_5$ has a layered crystal structure with space group P6/mmm and hexagonal lattice constants a = 5.5 Å and c = 9.3 Å [32]. It consists of a V3Sb kagome lattice bounded above and below by Sb honeycomb lattices and Cs hexagonal networks (Fig. 1c). The corresponding three-dimensional (3D) Brillouin zone (BZ) and a projected two-dimensional (2D) BZ on the (001) surface are illustrated in Fig. 1d, with high-symmetry points indicated. Theoretical calculations show that the Z$_2$ topological band structures of CsV$_3$Sb$_5$ are mainly located near the $\bar{M}$ point of the BZ.

We first focus on the normal state (T>T$_{CDW}$) electronic structure of CsV$_3$Sb$_5$. ARPES measurements were performed at 200 K (Fig. 2), and the Fermi-Dirac function was divided out in order to gain information about the thermally populated states slightly above E$_F$. Photon energy-dependent measurements reveal the 2D nature of the band structure (Fig.2b), and we use the projected 2D BZ ($\bar{\Gamma}$, $\bar{K}$, $\bar{M}$) hereafter. The measured band structure (Fig. 2c) exhibits excellent overall agreement with density functional theory (DFT) calculations (Fig. 2d). In order to make a quantitative comparison, characteristic features of the bands are marked in both experimental and calculated results (C1-C4 in Fig. 2e-g). While the Dirac-like crossing point at $\bar{K}$ (marked as C1) is well-aligned between the experiment and calculation, the other Dirac-like crossing point between $\bar{\Gamma}$ and $\bar{K}$ (C2) is at a slightly lower energy in the measured band. The band tops of two hole-like bands are shown in the calculation near $\bar{M}$ and $\bar{K}$ above $E_F$ (marked as C3 and C4, respectively). In the experiment, they are also observed above $E_F$, but at slightly lower energies compared to those of the calculation (Fig. 2e,f). Due to different orbitals of the bands, the measured band top at C4 is more evident with linear vertically polarized light (Fig. 2g). We note that these energy differences cannot be reconciled by a rigid band shift, pointing to the existence of vanadium *d* orbital correlation effects which are not included in the DFT calculations. Nevertheless, this bulk band renormalization effectively lowers the energy of the region where the theoretical TSSs should exist (Fig. 2e). Thus, we turn to the electronic bands along the $\bar{\Gamma}$ - $\bar{M}$ direction, where the TSSs are more separated from the bulk bands in the DFT calculations (Fig. 2i,k). An isolated hole-like feature is identified at the $\bar{M}$ point (Fig. 2j,l), which is not an extension of any bulk band (Fig. 2j-l). The momentum location and dispersion of this feature are very similar to those of the calculated TSSs (Fig. 2k), but the energy position is slightly lower. The tail of the experimental TSSs feature is very close to, or even slightly below, $E_F$ (Fig. 2j,l).

We next turn to the electronic structure in the CDW phase (T<$T_{CDW}$). While the Dirac-like crossing point at $\bar{K}$ (C1) shows little change with temperature, the other characteristic points (C2-C3) move toward lower energies in the CDW state, as clearly shown in Fig. 3(a-d). A systematic study of the temperature evolution of the C2 point (Fig. 3g) reveals that the band structure change takes place suddenly when the system enters the CDW phase, consistent with a proposed first-order transition [33]. This correlation between the band reconstruction and CDW is also supported by our DFT calculations. Recent simulation results have demonstrated the stabilization of kagome breathing mode distortions in DFT [34]. In our study, supercells representing the in-plane CDW in $CsV_3Sb_5$ are considered to simulate the electronic structure in the CDW phase (Fig. 3e, see Supplementary Materials for details of the simulation). Compared to the calculated bands in the normal state (T>$T_{CDW}$, Fig. 3f), the C2 and C4 points in the CDW state are pushed down by the opening of a gap in the energy region above C4 (marked by the orange box in Fig. 3e). This gap, however, has a much smaller effect on the Dirac point at $\bar{K}$ (C1). These results match the observations in the ARPES measurements and well explain the band reconstruction. Most importantly, in the region near $\bar{M}$, where the TSSs reside, the measured bands also move down in energy in the CDW phase (e.g., compare the energy position of C3 below and above $T_{CDW}$). This is also supported by the DFT results, in which the CDW-related gap opening lowers the energy of bands below $E_F$, with the greatest band reconstruction at $\bar{M}$ (Fig. 3e). A careful examination along the $\bar{\Gamma}$ - $\bar{M}$ cut reveals the expected hole-like TSSs at $\bar{M}$ with most of their density of states just below $E_F$ (Fig. 3h), consistent with the overall downward shift of the bands around $\bar{M}$.

After establishing the electronic structure associated with topological states and charge order in $CsV_3Sb_5$, we next search for the characteristic flat bands of the kagome lattice. As shown in Fig. 4a and b, a flat band around a binding energy ($E_B$) of -1.2 eV is seen across the entire momentum space. We note that a flat band feature also appears in our DFT calculations (Fig. 4c) at the same binding energy, which matches the experiment very well. Surprisingly, another nondispersive feature is also seen near $E_F$ (Fig. 4a,b,d). This feature is more pronounced when probed with linear vertically (LV) polarized light (Fig. 4d, see Supplementary Materials for details). We note that this nondispersive feature is gapped away from the $E_F$, as shown by the symmetrized image in Fig. 4d. Some other low energy bands can also disperse across this nondispersive band (Fig. 4a,b,d). Therefore, this feature is not related to the Fermi-cutoff, but represents a genuine nondispersive excitation in this material system.

Finally, we discuss the implications of our results. A unique cooperation is found between topology and the CDW in $CsV_3Sb_5$, which gives rise to the TSSs just below $E_F$. The energy modification induced by CDW is subtle but plays a key role in activating the topological physics in the superconducting state of $CsV_3Sb_5$, because only the electronic states below $E_F$ can contribute to transport at such a low temperature. The observation of a flat band similar to that of the DFT results confirms that the electron correlations effect induced by the kagome lattice are at play in the system. On the other hand, the observation of a nondispersive feature near $E_F$ is striking. While flat bands are naturally expected in kagome lattices, this near-$E_F$ feature is absent in our DFT calculations. We note that the energy location of a flat band in the kagome lattice is very sensitive to the hopping parameters (see Supplementary Materials), and not all the reported flat bands in real materials are captured by initial DFT calculations [28,29]. To what extent this discrepancy can be reconciled with the excellent overall agreement between the measured band structure and the DFT calculations, remains to be understood. In the context of kagome flat bands, the polarization dependence of the nondispersive feature is similar to that of a near-$E_F$ flat band reported in the kagome metal FeSn [29]. The energy position of this feature is also among the shallowest in all the reported flat bands in kagome metals. Other scenarios beyond kagome physics might involve the possible existence of quasi-one-dimensional stripe orders [22,23], a magic twist between adjacent layers [35-37], or the formation of dangling bonds on the surface [38,39], all of which could possibly give rise to some nondispersive features near $E_F$. Whether these scenarios are relevant to our observation remains to be explored. Regardless of the origin, the existence of a nondispersive band near $E_F$ would naturally give rise to a high electron density and bring strong correlations to the low-energy electrons which also participate in the formation of superconductivity. It would be interesting to investigate whether such strong correlations are related to the possible anomalous superconducting state in this system [16,24]. Our findings reveal that the charge order, topological states, electron correlation and superconductivity in $CsV_3Sb_5$ not only coexist but can also cooperate in a delicate fashion, which would open a new avenue for the realization of exotic quantum phases in kagome lattices.

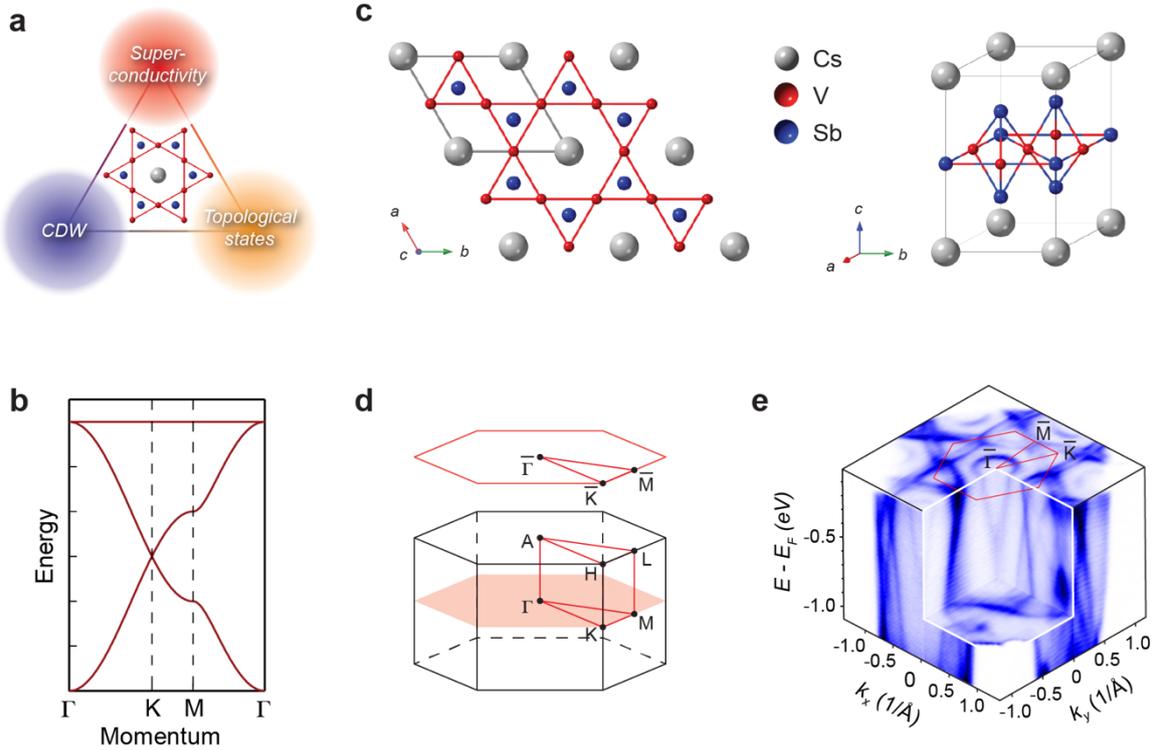

**Fig. 1| Crystal structure and electronic structure of the kagome superconductors CsV$_3$Sb$_5$. a** Exotic electronic orders in CsV$_3$Sb$_5$, including superconductivity, charge density wave (CDW) and topological states. **b** Tight-binding band structure of kagome lattice featuring a Dirac dispersion and a flat band. **c** Top view of the kagome plane (left) and the structural unit cell (right) of CsV$_3$Sb$_5$. **d** Schematic of the bulk and surface Brillouin zones along the (001) surface of CsV$_3$Sb$_5$. The high-symmetry points are marked. **e** 3D photoemission intensity plot measured with 90 eV photons.

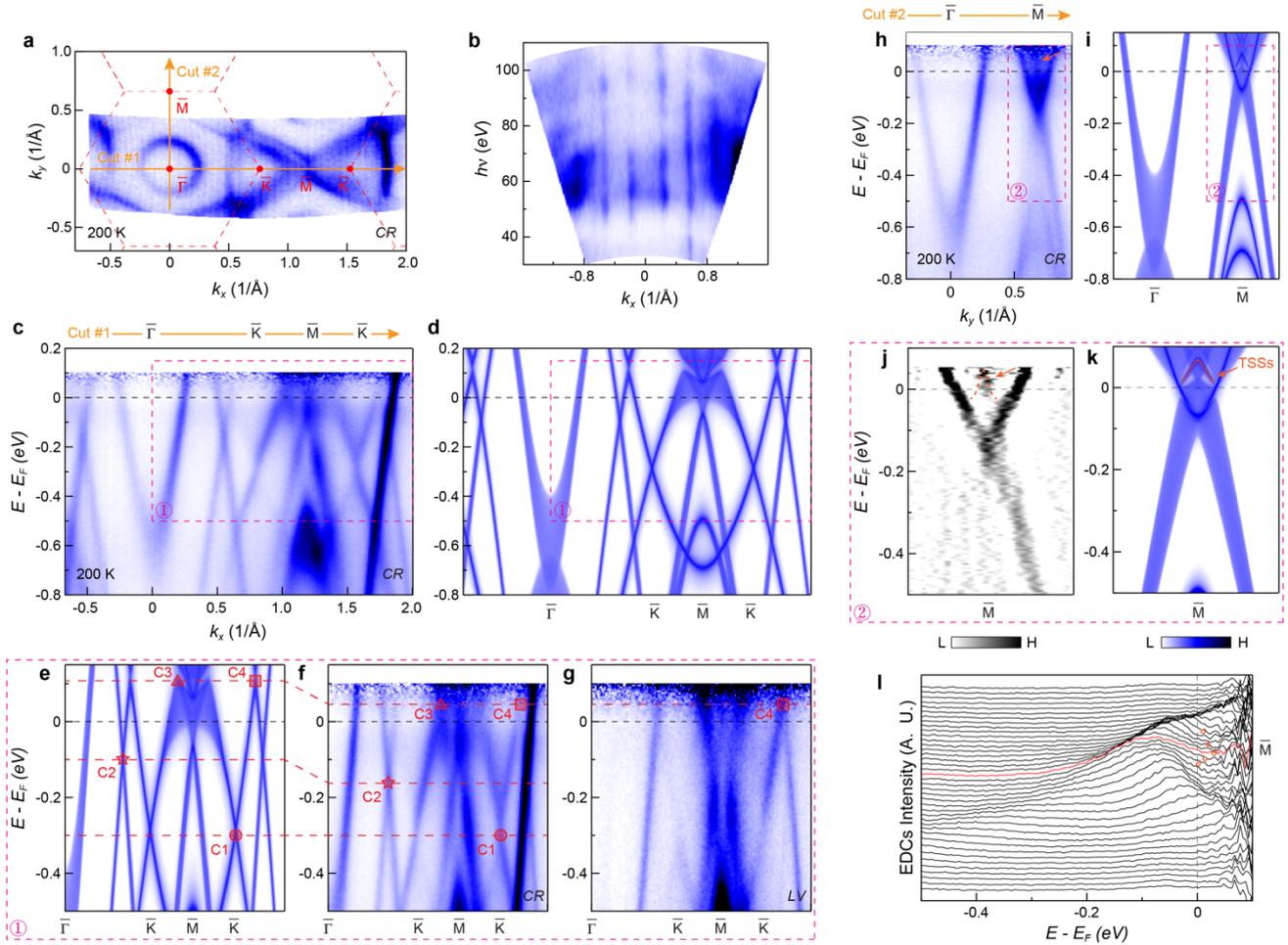

**Fig. 2 | Band structure above the CDW transition temperature. a** Fermi surface map of CsV$_3$Sb$_5$. **b** Photon energy-dependent ARPES spectral intensity map at the E$_F$ along the $\bar{K}$ - $\bar{\Gamma}$ - $\bar{K}$ direction. **c,d** Experimental (**c**) and calculated (**d**) band dispersion along the $\bar{\Gamma}$ - $\bar{K}$ - $\bar{M}$ - $\bar{K}$ direction. **e** Zoom-in plots of the calculated bands in a selected region (region 1) in (**d**), as indicated by the pink dashed box. **f,g** Experimental bands in the same region, measured with circular right (CR) and linear vertical (LV) polarization, respectively. The characteristic points (C1-C4) are marked with different markers. **h,i** Same as (**c,d**), but along the $\bar{\Gamma}$ - $\bar{M}$ direction. **j** Second derivative image of the original data with respect to momentum, for the selected region (region 2) in (**h**), as indicated by the pink dashed box. **k** Calculated bands in the same region. The orange dashed curve in (**j**) is an eye-guide for the TSSs. **l** Energy distribution curves (EDCs) around $\bar{M}$ point of the $\bar{\Gamma}$ - $\bar{M}$ momentum cut. Orange circles mark the energy features of the TSSs. All measurements were performed at 200 K, and the Fermi-Dirac function is divided out to reveal the energy region slightly above the E$_F$.

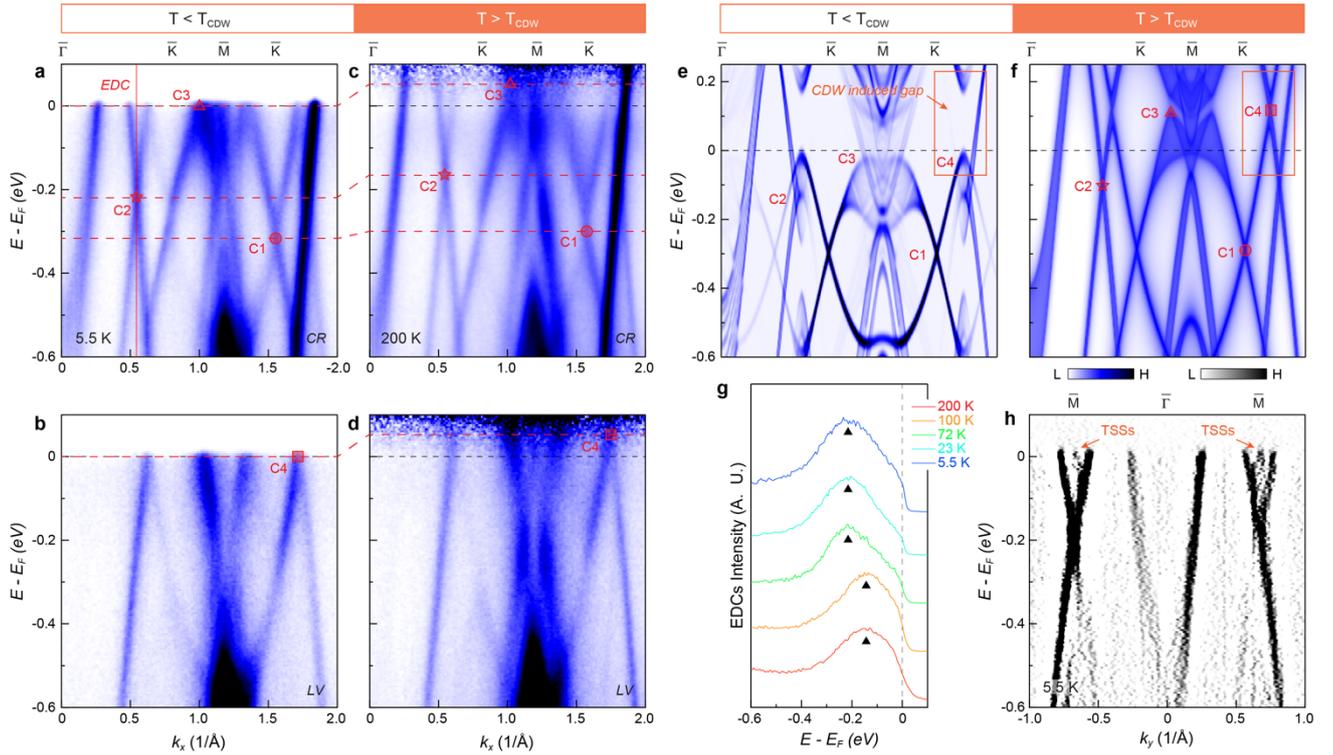

**Fig. 3 | Band structure below the CDW transition temperature. a,b** Experimental band dispersion in the CDW phase along the $\bar{\Gamma}$ - $\bar{K}$ - $\bar{M}$ - $\bar{K}$ direction collected at 5.5 K, with the CR and LV polarizations, respectively. The red circles, stars, triangles, and squares indicate the characteristic points C1, C2, C3, and C4, respectively. **c,d** Same as (**a,b**), but measured at 200 K. **e,f** Calculated bands along the $\bar{\Gamma}$ - $\bar{K}$ - $\bar{M}$ - $\bar{K}$ direction in the CDW phase with the Hex-Triangle structure and in the normal state (T>T$_{CDW}$), respectively. **g** Temperature evolution of the EDC at the Dirac-like crossing point C2, as indicated by the red line in (**a**). **h** Second derivative image with respect to momentum for the band structure along the $\bar{M}$ - $\bar{\Gamma}$ - $\bar{M}$ direction taken at 5.5 K.

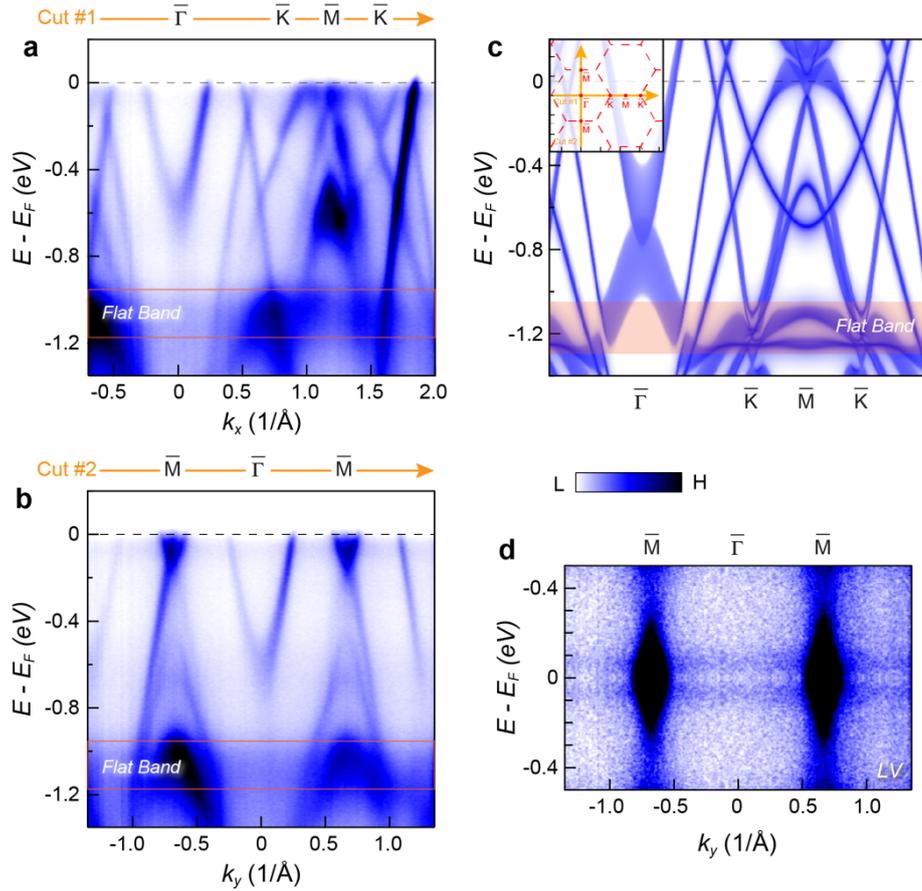

**Fig. 4| Identification of flat bands in CsV$_3$Sb$_5$. a,b** Experimental band dispersion along the $\overline{\Gamma}$ - $\overline{K}$ - $\overline{M}$ - $\overline{K}$ and $\overline{M}$ - $\overline{\Gamma}$ - $\overline{M}$ high symmetry directions, respectively. The orange box is an eye-guide for the nearly flat band. The bands are probed with circularly polarized light. **c** Calculated band structure along the $\overline{\Gamma}$ - $\overline{K}$ - $\overline{M}$ - $\overline{K}$ direction. Schematic of the locations of the momentum cuts is shown in the inset. **d** Symmetrized image of the band structure along the M - $\overline{\Gamma}$ - $\overline{M}$ direction, probed with linear vertically polarized light.